\documentclass[journal=iecred,manuscript=article]{achemso}
\usepackage{fancyhdr}
\usepackage{booktabs}
\usepackage{siunitx}
\usepackage{graphicx}
\usepackage{subfigure} 
\usepackage{float} 
\usepackage{tabularx}
\usepackage{color}
\usepackage{geometry}
\usepackage{natbib}
\usepackage{mciteplus}
\usepackage{footnote}
\usepackage{multirow}
\usepackage{color,soul}
\usepackage{longtable}

\usepackage{lineno}
\usepackage{multirow}
\usepackage{amssymb}
\usepackage{amsmath}
\usepackage{mathtools}
\usepackage{ragged2e}
\usepackage{blindtext}
\usepackage{commath}
\SectionNumbersOn
\usepackage[version=3]{mhchem}

\usepackage[labelsep=period]{caption}
\usepackage{titlesec}
\titlelabel{\thetitle.\quad}
\usepackage{longtable,booktabs,amsmath} 

\author{Yiyi Huo}

\affiliation{Department of Chemical and Materials Engineering, University of Alberta, Alberta T6G 1H9, Canada}
\alsoaffiliation{School of Chemical and Environmental Engineering, China University of Mining and Technology (Beijing), Beijing 100083, China}
\author{Mohammadhossein Golchin}
\affiliation{Department of Chemical and Materials Engineering, University of Alberta, Alberta T6G 1H9, Canada}
\author{Kaiyu Zhou}
\affiliation{Department of Chemical and Materials Engineering, University of Alberta, Alberta T6G 1H9, Canada}
\author{Ashwin Abraham}
\affiliation{Department of Chemical and Materials Engineering, University of Alberta, Alberta T6G 1H9, Canada}
\author{Somasekhara Goud Sontti}
\email{somasekhar.sonti@iitdh.ac.in}

\affiliation{Multiphase Flow and Microfluidics (MFM) Laboratory, Department of Chemical Engineering, Indian Institute of Technology Dharwad, Dharwad, 580011, Karnataka, India}

\author{Xuehua Zhang}
\email{xuehua.zhang@ualberta.ca}
\affiliation{Department of Chemical and Materials Engineering, University of Alberta, Alberta T6G 1H9, Canada}

\title[An \textsf{achemso} demo]
{Effects of Coal Particles on Microbubble-Enhanced Bitumen Separation in the Concentrated Slurry Flow of Oil Sands Tailings}

\begin{document}

\begin{tocentry}

\begin{figure}[H]
\centering
\includegraphics[width=65mm]{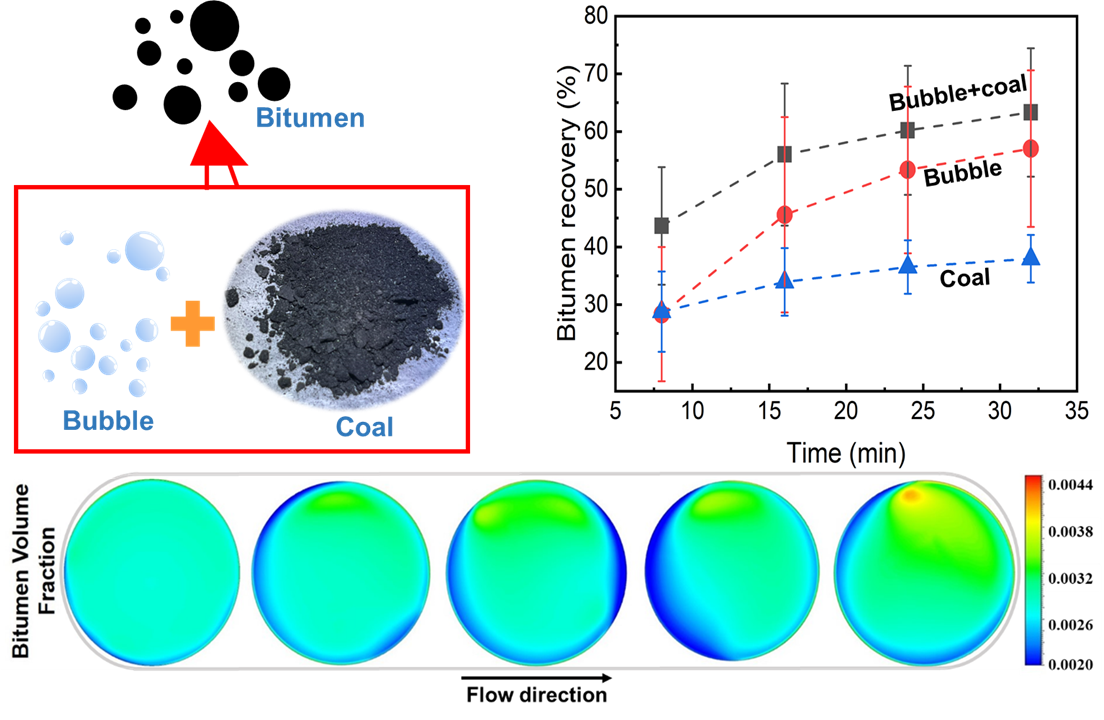}
\renewcommand{\captionfont}{\linespread{1.6}\normalsize}
\caption*{}
\label{TOC}
\end{figure}

\end{tocentry}
\newpage
\begin{abstract}

\noindent Our study investigates the segregation of bitumen residues within the transport pipeline before disposal in the presence of coal particles in carriers and microbubbles. Coal particles decreased the bitumen recovery by 17 \% without the injection of microbubbles. In addition, the improvement in bitumen recovery efficiency by 6 $mL$ $H_2O_2$ is negilible due to a small number of bubbles formed from $H_2O_2$ decomposition in the flow. However, the tremendous enhancement in the recovery efficiency was achieved with the simultaneous addition of coal particles and $H_2O_2$. Further increased recovery was noted as a higher volume of $H_2O_2$ was injected to form more microbubbles. Computational fluid dynamics (CFD) simulations were conducted to help understand effects of coal particles and microbubbles. The simulation results illustrated that the introduction of coal particles caused bitumen contents to accumulate in the middle of the pipe. Furthermore, an increased volume fraction of microbubbles contributed to a higher distribution of bitumen at the top of the pipe. This study not only offers valuable insights for developing an innovative strategy to enhance the efficiency of bitumen separation in hydrotransport processes but also contributes to a deeper understanding of the intricate interactions among bubbles, bitumen, and coal particles in a slurry flow.


\end{abstract}

\newpage
\section{INTRODUCTION}

 The warm water extraction is an established process to liberate bitumen  from oil sands mixtures by open-pit mining\cite{Sekhar2023POF,SCFD,masliyah2004understanding}. Through primary separation vessels, bitumen droplets are aerated and separated into froth, which floats at the top of the vessel. Nevertheless, residual bitumen still remains in the oil sands tailings and is transported to the tailing ponds via pipelines. Water-soluble organics, bitumen, and asphaltene attached to clay surfaces agglomerate small clay particles and cause water retention\cite{sontti2023numericals}. Reclamation of land and recycling of water are hindered in these stable suspensions.  Therefore, the removal of bitumen residues is crucial for pursuing sustainable development.
 
 The bubble flotation process is a widely used technique for separating substances in various industries, such as mining, wastewater treatment, and food processing \cite{zhou2022microbubble,zhang2022micro,scales}. A wide range of ultrasmall bubble generation techniques has been developed recently, including mechanical agitation, acoustic cavitation, and chemical reactions \cite{zhou2022microbubble,xing2017recent}. Since ultrasmall bubbles have high surface area-to-volume ratios, high internal pressure and along stability from pinning effects from the particle surface \cite{lohse2015surface,zhu2018diffusive,zhang2014perspectives}, they have unique physical and chemical properties, thus resulting in higher collision probabilities, higher separation efficiencies, and lower energy consumption.

The flotation kinetics are proportional to the particle size. The flotation recovery of fine/ultrafine particles is found to increase with particle size in multiple particle-bubble collision models \cite{zhou2020novel}. Bitumen droplets of higher size were found to be more efficient at recovering bitumen from high-grade ores\cite{zhou2000application}. Since residual bitumen droplets with small sizes and quantities are less likely to collide with bubbles, flotation treatment is difficult to recover fully.

A modified process for reducing the fine/ultrafine particle size before performing flotation is of great significance for further improving recovery efficiency. A number of modified processes are employed, including selective flocculation, agglomerate flotation, electro-coagulation, oil flotation, and carrier flotation\cite{zhou2020novel}.
A carrier flotation process is a type of hydrophobic flocculation flotation. In order to collect fine hydrophobic particles, it utilizes large hydrophobic carrier particles controlled by chemical and hydrodynamic conditions. \citet{chia1983theoretical} indicates that stirring vigorously causes anatase fine particles to aggregate with calcite larger particles. Because enlarged particles are more likely to attach to bubbles, larger anatase particles performed better in flotation.

 The typical hydrodynamic cavitation and the use of $H_2O_2$ have several difference. $H_2O_2$ can generate bubbles from chemical decomposition.  Moreover, the hydrodynamic flow in the pipeline enhanced the mixing and transfer of $H_2O_2$. The utilization of $H_2O_2$ does not require intensified energy input. In contrast, for hydrodynamic cavitation, the solid contents may wear venturi tube in use. To generate microbubbles by hydrodynamic cavitation for bubble formation, the combination of slurry flow rate, gas flow rate and slurry compositions requires optimization adjustments. The effects of the oxygen bubble production from $H_2O_2$ decomposition offers a simple, low-energy method for bubble enhanced separation. 
 
 Recent studies have reported the application of microbubbles in extracting residual bitumen from oil sands tailings in pipeline transportation. Microbubbles were generated by hydrodynamic cavitation in a cavitation tube or by hydrogen peroxide ($H_2O_2$) decomposition \cite{zhou2023enhanced,zhou2022microbubble}. From the measurements by using lab-scale pipeline loop, microbubbles formed from $H_2O_2$ decomposition could help to achieve an optimal recovery efficiency of 70 \% from artefaical tailings \cite{zhou2023enhanced}. 

Coal is a kind of naturally hydrophobic carbonaceous mineral\cite{gutierrez-rodriguezEstimatingHydrophobicityCoal1984}. The hydrophobicity of a coal particle is determined by the chemical composition of its surface aromatic and aliphatic groups \cite{dingInvestigationBituminousCoal2009}. Froth flotation is a method that enhances the separation of particles by attaching to bubbles for a lower density \cite{polatPhysicalChemicalInteractions2003}. It was reported that the addition of bitumen droplets increased the efficiency of carrier flotation \cite{atesokCarrierFlotationDesulfurization2001}. The research work conducted by \citet{zhou2013role} reveals that using coal particles (150 $\mu$m) as carriers for flotation increased bitumen recovery by 20-30 \%. The coal minerals can also be ground to a specific size and are readily accessible.

The present study introduces an a innovative techniques such as utilizing coal particles and microbubbles for bitumen removal residual in a hydrotransport pipeline. We attempted to address a critical need in the industry for more efficient and environmentally friendly processes. In this work, the bitumen separation from concentrated tailings was promoted through microbubble injection using fine coal particles as carriers.  So far few studies have focused on using hydrophobic particles as carriers and microbubbles to recover residual bitumen coal particles are sometimes found in oil sands mining sites. To study the effects of microbubbles and coal particle addition on bitumen recovery, we combine experiments and computational fluid dynamics~(CFD) simulations. Experimental and simulation studies show that the synergistic effect of microbubbles and coal particles enhances bitumen recovery. Such pronounced synergistic effect may be explored in the separation process where small hydrophobic particles (such as wood debris or micro plastics) are mixed with bubbles in the slurries.

\section{MATERIALS AND METHODOLOGY}
\subsection{Laboratory-Scale Pipeline Loop}
A laboratory-scale hydrotransport pipeline test loop simulates conditions in pipelines carrying oil sands tailings as shown in Figure.~S1. In this experimental setup, microbubble injection is applied to recover residual bitumen from pipeline tailings. The hydrotransport lab-scale pipeline loop is composed of a stainless steel pipe with an inner diameter of 20.64 $mm$ and a length of approximately 4 $m$. Figure.~S1 shows a schematic sketch of the loop. A progress cavity pump, an open collection cell, a view window, a heat exchanger, and a drain valve are some of the primary components as shown in Figure.~S1. Tailings are circulated in the loop at a velocity of 2 $m/s$ using a low-speed progressive cavity pump powered by a variable AC power supply of 3 horsepower (HP). The temperature of the fluid is controlled by a double-pipe heat exchanger connected to a programmable circulation bath. The cell serves as both a location to insert the feed and collect the floated froth. The cell can also be used to view the appearance of recovered froth. 
Glass view windows measure 40 $cm$ in length and 17 $mm$ in diameter, enabling direct visual and recorded tracking of tailing slurries while in motion. The ball valve drains waste fluid at the end of the experimental process.

\subsection{Preparation of Artificial Oil Sands Tailings}

The artificial oil sands tailings comprise a mixture of dry sands, coal particles, bitumen, and process water.
In this study, coking coal from western Canada was used. The coal samples were ground into fine coal particles smaller than 150 $\mu$m. The large coal lumps were crushed into smaller pieces and then ground into tiny particles by a ball mill. To obtain the desired particle size, the ground coal particles were sieved through 110 mesh. 

The preparation of artificial oil sands tailings started by partially hydrating the sands (Sil Industrial Minerals) with 7 \% of the total process water content provided by Imperial Oil Limited. The ground coal particles and pre-heated bitumen (80 \textdegree C) were mixed with hydrated sands thoroughly, followed by the addition of rest process water and heavy stirring. A control without adding coal particles were carried out.
The properties of artificial oil sands tailings are shown in Table~\ref{tab1}. The used process water in each experiment is consistent. Thus, the concentration of chemistry compositions did not influence the results. The compositions of artificial oil sands tailings with the addition of coal and bubbles in this study are shown in Table~\ref{tab2}.

\begin{table}[!ht]
\centering
\caption{Properties of the prepared artificial tailings.}
\renewcommand{\arraystretch}{1.5}  
\label{tab1}
\begin{tabular}{|c|c|}
\hline
compositions            & parameter                         \\ \hline
\text{process water} & \text{}                          \\
pH \cite{gao2021formation}                    & 7.5                                \\
surface tension \cite{gao2021formation}        & 70.1 mN/m                          \\
chemistry composition \cite{gao2021formation}  & $NH_4^+$,$Ca^{2+}$,$Cl^-$,$Na^+$,$Mg^{2+}$,$SO_4^{2-}$,$HCO_3^-$ \\ \hline
\text{coal}          & \textbf{}                          \\
size                   & \textless 150 µm                   \\ \hline
\text{sands}         & \textbf{}                          \\
size                   & \textless 200 mesh                 \\
mineral composition & $SiO_2 \approx 93.8\%;~ Al_{2}O_{3} < 4.00\%;~ Fe_{2}O_{3} < 1.20\%;~ TiO_2 < 0.01\%$ \\ \hline

\end{tabular}
\end{table}
\vspace{0.3cm}
\begin{table}[!ht]
\centering
\caption{Compositions of artificial oil sands tailings with the addition of coal and bubbles. For all cases, bitumen, sands, and water mass percentages(wt$\%$) are respectively 0.2, 50, and 49.8.}
\label{tab2}
\begin{tabular}{|c|c|c|}
\hline
 Experimental case & coal (g) & bubble ($mL$/$H_2O_2$) \\  \hline
case a      & 0        & 0     \\
case b      & 0        & 6     \\
case c      & 40       & 0      \\
case d      & 40       & 6      \\
case e      & 40       & 12     \\\hline                         
\end{tabular}
\end{table}

\subsection{Bitumen Separation and Quantification of Recovered Froth}
Initially, the artificial oil sands tailings were transferred into the loop through a trough. The temperature and flow velocity of the tailings were respectively kept at 42 \textdegree C and 2 $m/s$. Microbubbles formation is implemented by injecting $H_2O_2$ into the tailings at a fixed flow rate of 0.5 $mL/min$ for 12 and 24 minutes separately.
Collection of the recovered froth was carried out the trough for every 8 minutes and lasted for 32 minutes. The trough with $28~\mathrm{~cm}$ height, $20~\mathrm{~cm}$ length, and $5~\mathrm{~cm}$ width allows the aerated bitumen to move upward by microbubbles and form a bitumen froth layer at the top of the trough. A spatula was used to transfer the froth into thimbles. 

Afterwards, a Dean-Stark analysis was performed to ascertain the composition of the froth in the thimbles. Toluene washed out the bitumen from the froth and the contained water was first evaporated at 250 \textdegree C and collected in the vial by cooling effect. The solid contents were remained in the thimble. 5 $mL$ of homogeneous toluene solution containing bitumen was extracted and dried on the filter paper. Afterwards, the bitumen content in 5 $mL$ of the solution was weighted. In order to obtain the recovered coal content, the remaining solids in the thimble were dried and combusted in the furnace. The coal content can be calculated according to the mass loss of the dry solids after combustion.

\subsection{NUMERICAL SIMULATION}

\subsubsection{Multi-Fluid Model}

An Eulerian-Eulerian (E-E) multifluid model (MFM) represents fluid systems as interpenetrating continua, treating distinct phases separately within the E-E model. In E-E MFM, phase coupling is achieved through pressure and exchange coefficients, playing significant roles. Interphase exchange coefficients and pressures govern phase coupling, and the model considers energy dissipation, exchange between particles, and interfacial forces~\cite{li2018hydrodynamic,pressuredropocean,sontti2023numericals,SCFD,Sekhar2023POF}. The continuity equation (Eqs. 1(a) and 1(b)) is applied to all phases within the multiphase system. The primary focus is on the liquid phase momentum balance, acting as the carrier fluid. Interphase forces, represented by $\Vec{F}_{l,s}$ and $\Vec{F}_{l,b}$, are considered for interactions between the liquid and solids, as well as the bitumen phases, respectively~\cite{SCFD,Sekhar2023POF}. The model integrates interactions between the liquid and solid phases, as shown in Table 3.

The drag force, crucial in modeling multiphase slurry flows, is expressed in Eqs. 2(b) to 2(j), accounting for the difference between the primary phase velocity and secondary phase velocity. The Gidaspow drag model~\cite{Gidaspow} accurately describes the drag force between carrier phase and solid phase. In turbulent slurry flows, the flow behavior is significantly influenced by the turbulent dispersion force arising from interactions between turbulent eddies and secondary phases. The CFD model incorporates the turbulent dispersion force, expressed in Eqs. 2(d), 2(g), and 2(j), utilizing the approach introduced by ~\citet{Burnetal} This formulation accounts for interactions between the carrier fluid and solid particles.

The CFD model takes into account the interactions between carrier fluid and bitumen droplets by considering drag  forces (Eq. 2(f) and 2(i), which are similar to the treatment of carrier-solids interactions. The E-E MFM model is not inherently compatible with the Derjaguin, Landau, Verwey, and Overbeek (DLVO) theory. The DLVO interactions dominate particle aggregation, dispersion, and deposition. However, in the present study the interactions between particles and bubbles are ignored because of the complexity of the system. The solid phase, coal, and bitumen phase momentum equations are listed from Eqs. 3(a) to 5(d).  The symmetric drag model~\cite{Sekhar2023POF} is employed to describe the interactions between the carrier fluid and bitumen droplets, as well as the interactions between the carrier fluid and bubbles, and between solids (coal) and bitumen droplets (bubbles). The turbulent dispersion force is modeled following the methodology by Burns et al.~\cite{Burnetal}. Additionally, the drag force proposed by Gidaspow et al.~\cite{Slurryocean,Burnetal,Gidaspow} is employed to represent solid-coal interphase forces. The system operates at a Reynolds number of 2400. At the outlet, the boundary condition is specified as gauge pressure, while at the inlet, a constant velocity of 2 $m/s$ is maintained.



\begin{table}[!ht]
\caption{Momentum equations.~\cite{SCFD,ParvaEulerianmultifluid,Sekhar2023POF}}
\label{tab:momentum_eq}

\renewcommand{\arraystretch}{1.2}
\setlength{\tabcolsep}{4pt} 
\centering
\begin{tabular}{l l}
\hline
\textbf{Continuity} & \\[2pt]
\hline
\multicolumn{2}{l}{
$\displaystyle \frac{\partial}{\partial t}(\rho_q\alpha_q) + \nabla \cdot (\rho_q\alpha_q \Vec{v}_q) = 0, \quad (q = l,s,c,b)$ \quad (1a)
} \\
\multicolumn{2}{l}{
$\displaystyle \alpha_l + \alpha_{s_1} + \alpha_{s_2} + \alpha_{c_1} + \alpha_{c_2} + \alpha_b = 1$ \quad (1b)
} \\
\hline
\textbf{Momentum (carrier fluid)} & \\[1pt]
\hline
\multicolumn{2}{l}{
$\displaystyle
\frac{\partial}{\partial t}(\rho_l\alpha_l \Vec{v}_l) + \nabla \cdot (\rho_l\alpha_l \Vec{v}_l \otimes \Vec{v}_l)
= -\alpha_l \nabla P + \rho_l \alpha_l \Vec{g} + \nabla \cdot \tau_l + \Vec{F}_{1,_s} + \Vec{F}_{1,_c} + \Vec{F}_{1,_b} 
$ \quad (2a)
} \\
\multicolumn{2}{l}{
$\displaystyle
\Vec{F}_{l,_{s_i}} = \Vec{F}_{td,l_{s_i}} + \Vec{F}_{drag,l_{s_i}}, \quad
\Vec{F}_{drag,l_{s_i}} = \frac{3}{4} C_D \alpha_{si}\rho_l \frac{||\Vec{v}_l-\Vec{v}_{si}||}{d_p} (\Vec{v}_l-\Vec{v}_{si}),
$ \quad (2b, 2c)
} \\
\multicolumn{2}{l}{
$\displaystyle
\Vec{F}_{td,l_{s_i}} = \frac{3}{4}\frac{C_D \mu_{t,l}}{d_p \sigma_{t,l}}\alpha_{si} 
||\Vec{v}_l-\Vec{v}_{si}|| \left( \frac{\nabla \alpha_{si}}{\alpha_{si}} - \frac{\nabla \alpha_l}{\alpha_l} \right)
$ \quad (2d)
} \\
\multicolumn{2}{l}{
$\displaystyle
\Vec{F}_{l,_{c_i}} = \Vec{F}_{td,l_{c_i}} + \Vec{F}_{drag,l_{c_i}}, \quad
\Vec{F}_{drag,l_{c_i}} = \frac{3}{4} C_D \alpha_{ci}\rho_l \frac{||\Vec{v}_l-\Vec{v}_{ci}||}{d_p} (\Vec{v}_l-\Vec{v}_{ci})
$ \quad (2e, 2f)
} \\
\multicolumn{2}{l}{
$\displaystyle
\Vec{F}_{td,l_{c_i}} = \frac{3}{4}\frac{C_D \mu_{t,l}}{d_p \sigma_{t,l}}\alpha_{ci} 
||\Vec{v}_l-\Vec{v}_{ci}|| \left( \frac{\nabla \alpha_{ci}}{\alpha_{ci}} - \frac{\nabla \alpha_l}{\alpha_l} \right)
$ \quad (2g)
} \\
\hline
\textbf{Momentum (solid phase)} & \\[1pt]
\hline
\multicolumn{2}{l}{
$\displaystyle
\frac{\partial}{\partial t}(\rho_{si}\alpha_{si} \Vec{v}_{si}) + \nabla \cdot (\rho_{si}\alpha_{si} \Vec{v}_{si} \otimes \Vec{v}_{si})
= -\alpha_{si} \nabla P - \nabla P_{si} + \rho_l \alpha_{si} \Vec{g} + \nabla \cdot \tau_{si} + \Vec{F}_{{si},l} 
$ \quad (3a)
} \\
\multicolumn{2}{l}{
$\displaystyle
\beta_{ij} = \frac{3(1+e_{ij})\left(\frac{\pi}{2} + C_{fr,ij}\frac{\pi^2}{8}\right)
\alpha_{si}\rho_{si} \alpha_{sj}\rho_{sj} (d_{si}+d_{sj})^2 g_{0,ij}}
{2\pi (\rho_{si} d^3_{si} + \rho_{sj} d^3_{sj})} ||\Vec{v}_{si}-\Vec{v}_{sj}|| 
$ \quad (3b)
} \\
\multicolumn{2}{l}{
$\displaystyle
\tau_{si} = \alpha_{si}\mu_{si} (\nabla \Vec{v}_{si} + (\nabla \Vec{v}_{si})^T) 
+ \alpha_{si} \left(\lambda_{si}-\frac{2}{3}\mu_{si}\right) (\nabla \cdot \Vec{v}_{si}) \Bar{\Bar{I}} 
$ \quad (3c)
} \\
\hline
\end{tabular}
\end{table}

\begin{table}[!ht]
\renewcommand\thetable{3~~(continued)}
\caption{Momentum equations.~\cite{SCFD,ParvaEulerianmultifluid,Sekhar2023POF}}
\label{tab:Eq145}
\renewcommand{\arraystretch}{1.2}
\centering
\begin{minipage}{\linewidth}

\textbf{Momentum (\textit{i}th coal phase)} \\[4pt]

\begin{align}
\begin{split}
\frac{\partial}{\partial t}(\rho_{ci}\alpha_{ci}\Vec{v}_{ci}) 
+ \nabla \cdot (\rho_{ci}\alpha_{ci}\Vec{v}_{ci} \otimes \Vec{v}_{ci})
&= -\alpha_{ci}\nabla P - \nabla P_{ci} + \rho_{l}\alpha_{ci}\Vec{g}  \\
&\quad + \nabla \cdot \tau_{ci} + \Vec{F}_{{ci},l}
+ \Vec{F}_{drag,{{ci},s}} + \Vec{F}_{drag,{{ci},b}}\beta_{ij}(\Vec{v}_{cj}-\Vec{v}_{ci})
\end{split}
\tag{4a}
\end{align}

\begin{align}
\beta_{ij} = 
\frac{3(1+e_{ij})
\left(\frac{\pi}{2}+C_{fr,ij}\frac{\pi^2}{8}\right)
\alpha_{ci}\rho_{ci}\alpha_{cj}\rho_{cj}(d_{ci}+d_{cj})^2 g_{0,ij}}
{2\pi(\rho_{ci}d^3_{ci}+\rho_{cj}d^3_{cj})} 
\, \|\Vec{v}_{ci}-\Vec{v}_{cj}\|
\tag{4b}
\end{align}

\begin{align}
\tau_{ci} = 
\alpha_{ci}\mu_{ci}(\nabla\Vec{v}_{ci} + (\nabla\Vec{v}_{ci})^T)
+ \alpha_{ci}\left(\lambda_{ci}-\frac{2}{3}\mu_{ci}\right)
(\nabla \cdot \Vec{v}_{ci})\Bar{\Bar{I}}
\tag{4c}
\end{align}

\begin{align}
\Vec{F}_{drag,c} = -\Vec{F}_{drag,l}, 
\qquad
\Vec{F}_{td,c} = -\Vec{F}_{td,l}
\tag{4d–4e}
\end{align}

\textbf{Momentum (bitumen phase)} \\[4pt]

\begin{align}
\begin{split}
\frac{\partial}{\partial t}(\rho_{b}\alpha_{b}\Vec{v}_{b}) 
+ \nabla \cdot (\rho_{b}\alpha_{b}\Vec{v}_{b} \otimes \Vec{v}_{b})
&= -\alpha_{b}\nabla P + \rho_{l}\alpha_{b}\Vec{g}  \\
&\quad + \nabla \cdot [\,\mu_{b}(\nabla\Vec{V} + \nabla\Vec{V}^T)\,]
+ \Vec{F}_{{b},l} + \Vec{F}_{{b},si} + \Vec{F}_{{b},ci}
\end{split}
\tag{5a}
\end{align}

\begin{align}
\Vec{F}_{b,l} = -\Vec{F}_{l,b}, \quad
\Vec{F}_{b,si} = -\Vec{F}_{si,b}, \quad
\Vec{F}_{b,ci} = -\Vec{F}_{ci,b}
\tag{5b–5d}
\end{align}

\end{minipage}
\end{table}

\renewcommand{\arraystretch}{1.5}

\subsubsection{Turbulence Model}

The choice of a turbulence model has been a topic of discussion for various applications. The feasibility and implementation of turbulence models, such as the \textbf{$k-\epsilon$} and \textbf{$k-\omega$} models, have been extensively studied and reported in the context of slurry systems. However, recent research suggests that the standard \textbf{$k-\epsilon$} turbulence model is suitable to describe slurry systems~\cite{li2018hydrodynamic,Sekhar2023POF}. In this study, the \textbf{$k-\epsilon$} turbulence model, obtained through Reynolds-averaged Navier-Stokes (RANS) equations, is employed to accurately capture the turbulence in the slurry mixtures~\cite{puhan2023insights,sontti2023numericals}. The equations and constants for the \textbf{$k-\epsilon$} model can be found in Table~\ref{tab:Turbulence} from Eqs.~6(a)-6(f)~\cite{abdulbubble,slurry,cicepigging,launder}

 

The standard wall functions are used for handling wall effects, based on Launder-Spalding's law of the wall. For wall-bounded flows with high Reynolds numbers, these wall functions have demonstrated reasonably accurate predictions. Similar models have been used extensively for horizontal pipelines.~\cite{cGauravslurrypowder,kepsilonturb,Slurryocean,Rans23d,wliuhozslurry}. In the present study, the $\begin{alignedat}{8} y^{+} \end{alignedat}$ falls within the range of 30  $\begin{alignedat}{8} < y^{+} < \end{alignedat}$  300 for the carrier and secondary phases. This validates the logarithmic law for mean velocity within the system. The model employing standard wall functions demonstrates quicker and smoother convergence in contrast to the  \textit{SST} $\begin{alignedat}{8} k-\omega \end{alignedat}$ model, with minimal near-wall effects. As a result, the Standard Wall Functions treatment, has been adopted for this study.
\\
\begin{table}[!ht]
    \renewcommand\thetable{4}

    \centering
    \caption{Standard $\begin{alignedat}{8} k-\epsilon \end{alignedat}$  turbulence model.\cite{li2018hydrodynamic}}
\label{tab:Turbulence}

\begin{tabular}{l l l}
\hline
      the $\begin{alignedat}{8} k \end{alignedat}$  equation & $ \begin{alignedat}{8}\frac{\partial}{\partial t}(\rho_{m} K)+\nabla(\rho_{m}\Vec{v}_{m}K) =\nabla . (\frac{\mu_{t,m}}{\sigma_{k}}\nabla K) + G_{k,m}-\rho_{m}\epsilon\end{alignedat}$\ & (6a)\\
      the $\begin{alignedat}{8} \epsilon \end{alignedat}$ equation & $\begin{alignedat}{8}\frac{\partial}{\partial t}(\rho_{m}\epsilon)+\nabla(\rho_{m}\Vec{v}_{m}\epsilon) =\nabla . (\frac{\mu_{t,m}}{\sigma_{k}}\nabla\epsilon) +\frac{\epsilon}{K} (C_{1\epsilon}G_{k,m}-C_{2\epsilon}\rho_{m}\epsilon)\end{alignedat}$ & (6b)\\
      mixture density & $\begin{alignedat}{8}\rho_{m}=\Sigma_{q=1}^{n}\alpha_{q}\rho_{q}\end{alignedat}$ & (6c)\\
      mixture velocity & $\begin{alignedat}{8}\Vec{v}_{m}=\frac{\Sigma_{q=1}^{n}\alpha_{q}\rho_{q}\Vec{v}_{q}}{\Sigma_{q=1}^{n}\alpha_{q}\rho_{q}}\end{alignedat}$ & (6d)\\
      turbulent viscosity & $\begin{alignedat}{8}\mu_{t,m}=\rho_{m}C_{\mu}\frac{K^2}{\epsilon}\end{alignedat}$ & (6e)\\
      standard constants & $\begin{alignedat}{8}C_{1\epsilon}=1.44, C_{2\epsilon}=1.92, C_{\mu}=0.09, \sigma_{k}=1.0, \sigma_{\epsilon}=1.3\end{alignedat}$ & (6f)\\
      \hline
\end{tabular}
\end{table}

\newpage
\subsubsection{Carrier Fluid Rheological Model}
The non-Newtonian behavior of the carrier fluid is described by the Casson model~\cite{Adeyinka}. The Casson rheological model is suitable for fine sand and coal particles suspended in water, particularly in concentrations between 10\% and 40\%. This study employs the Casson model since the fines ($<$44 $\mu$m) weight fraction falls within this range. The Casson model describes the non-Newtonian behavior of the carrier, as defined by Eq. (7), and is coupled to the momentum equation (Eq. (2a)).

\begin{center}
\begin{equation}  \tag{7}\label{eqn:Cassonmodel}
	\tau_{{l}}^{1/2}=\tau_{{y}}^{1/2}+\mu_{{c}}^{1/2} \dot{\gamma}^{1/2}
\end{equation}  
\end{center}
\subsubsection{Computational Domain and Boundary Conditions}
A cylindrical pipeline with a diameter (D) of 2.2 $cm$ and a length (Z) of 2 $m$, is considered to represent lab-scale pipeline conditions. A transient E-E model is utilized, wherein distinct phases are treated as simultaneous continuous systems. Uniform velocity and volume concentration are assumed for each phase at the pipe inlet. The outlet boundary condition is set to atmospheric pressure. At the wall boundary the no-slip condition indicates that the carrier fluid velocity to be zero. The CFD model accounts for the flow conditions at both the entrance and exit of the pipeline. Additionally, the diameter of the pipeline is relatively small compared to its length, indicating a more uniform flow distribution along the length of the pipeline. These factors will contribute to minimizing entrance and end effects in the simulation, ensuring that the results are reliable and accurately represent the behavior of the flow in the pipeline. The kinetic theory of granular flow (KTGF) is utilized to describe particle interactions. All phases pressure is the same, while their corresponding conservation equations are solved separately. Continuous and dispersed phase volume fractions are assumed to be continuous functions of both space and time, collectively summing up to unity.  The E-E model is chosen for its comprehensiveness and computational robustness in representing secondary phase interactions. The specific E-E model (i.e., Multifluid Model) is employed in this study. This modeling approach can also be used in industrial stages with much larger diameters than lab-scale systems.\cite{sontti2023numericals} CFD E-E models have also been used to study a wide range of industrial problems by several researchers.~\cite{SCFD,elkarii2023global,chen2020study}\\

\subsubsection{Simulation Method}
The conservation equations for the multi-fluid model with KTGF are listed in Table S1 and Table S2. The finite volume method (FVM) based ANSYS Fluent solver 2022 R2 is used to solve the unsteady state equations. Table S3 provides the solver settings and schemes used in E-E MFM. The turbulence intensity and turbulent viscosity ratio for all phases are defined as 5\% and 10\%, respectively. A comprehensive compilation of fluid properties and KTGF model parameters is available in Table~\ref{tab:properties}. The second-order upwind method is employed to solve momentum equations and turbulence transport. The volume fraction is computed using the first-order upwind method, and specified relaxation factors are applied for pressure, momentum, and volume fraction. The transient simulations involve the discretization of governing equations in both space and time. Temporal discretization follows the same spatial discretization as steady-state equations, utilizing the second-order implicit scheme for stability~\cite{A.}.\\


 The convergence criterion of $\begin{alignedat}{8} 10^{-4} \end{alignedat}$ is set for the relative error between two successive iterations in terms of scaled residual components, which includes the governing equations encompass continuity and momentum components for the carrier fluid, solid phases, coal phases, bitumen, and bubbles. Additionally, the gravity $\begin{alignedat}{8} g = -9.8 \hspace{0.2cm} m^{2}/s   \end{alignedat}$   is in the opposite direction of Y-axis. Time step sensitivity analysis from 0.0001 $s$ to 0.001 $s$ reveals no significant changes in the results, thus a fixed time step of 0.001 $s$ is adopted for all cases in this study.  All computational calculations for this study are conducted at the Compute Canada Graham cluster, utilizing the high-performance computing (HPC) facility. The simulations are performed using 44 CPUs. The flow behavior is investigated over 28 $s$ flow time, corresponding to solving the equations for 28,000 time steps. To analyze the results, a cross-sectional XY plane is created, and the data are examined at Z = 1.98 $m$, representing a specific location along the pipeline. Distribution profiles were then analyzed based on this XY plane. Additionally, a vertical reference line was drawn in the middle of the XY plane to facilitate further analysis and comparison of results.
\\

\begin{table}[H]
    \renewcommand\thetable{5}

    \centering
    \caption{Boundary conditions and material properties used in CFD simulations.}
    \label{tab:properties}
 \renewcommand{\arraystretch}{1.1}
\begin{centering}
\begin{tabular}{l l }

\hline
parameter & value \\
\hline
pipe diameter(m) & 0.022  \\
pipe length (m) & 2\\
solid diameter ($\mu$m) & 60 and 100\\
density of the solid particles ($kg/m^3$)  &  2650\\
coal diameter ($\mu$m) & 50 and 120\\
density of the coal particles ($kg/m^3$)  &  1650\\
carrier density ($kg/m^3$) & 1219.7\\
bitumen droplet diameter($\mu$m) & 400 \\
density of the bitumen droplets ($kg/m^3$)  &  990\\
bitumen viscosity ($Pa$ s) & 20 \\
air bubble diameter($\mu$m) & 500 \\
density of the air bubbles ($kg/m^3$)  &  1.225\\
air viscosity ($Pa$ s) & 1.7e-5 \\
casson viscosity $\mu_c$ ($Pa$ s) &0.001802~\cite{Adeyinka}\\
yield stress, $\tau_y$ ($Pa$) & 0.0512~\cite{Adeyinka}\\
fraction packing limit & 0.60~\cite{A.}\\
angle of internal friction & 30~\cite{A.}\\
particle-particle restitution coefficient & 0.90 \cite{ppcollision}\\
particle-wall specularity coefficient & 0.2~\cite{pwallcollision}\\

\hline

\end{tabular}
\end{centering}
\end{table}

\subsubsection{Computational Mesh Study}

\begin{figure}[!ht] 
	\centering
	\includegraphics[width=0.8\columnwidth]{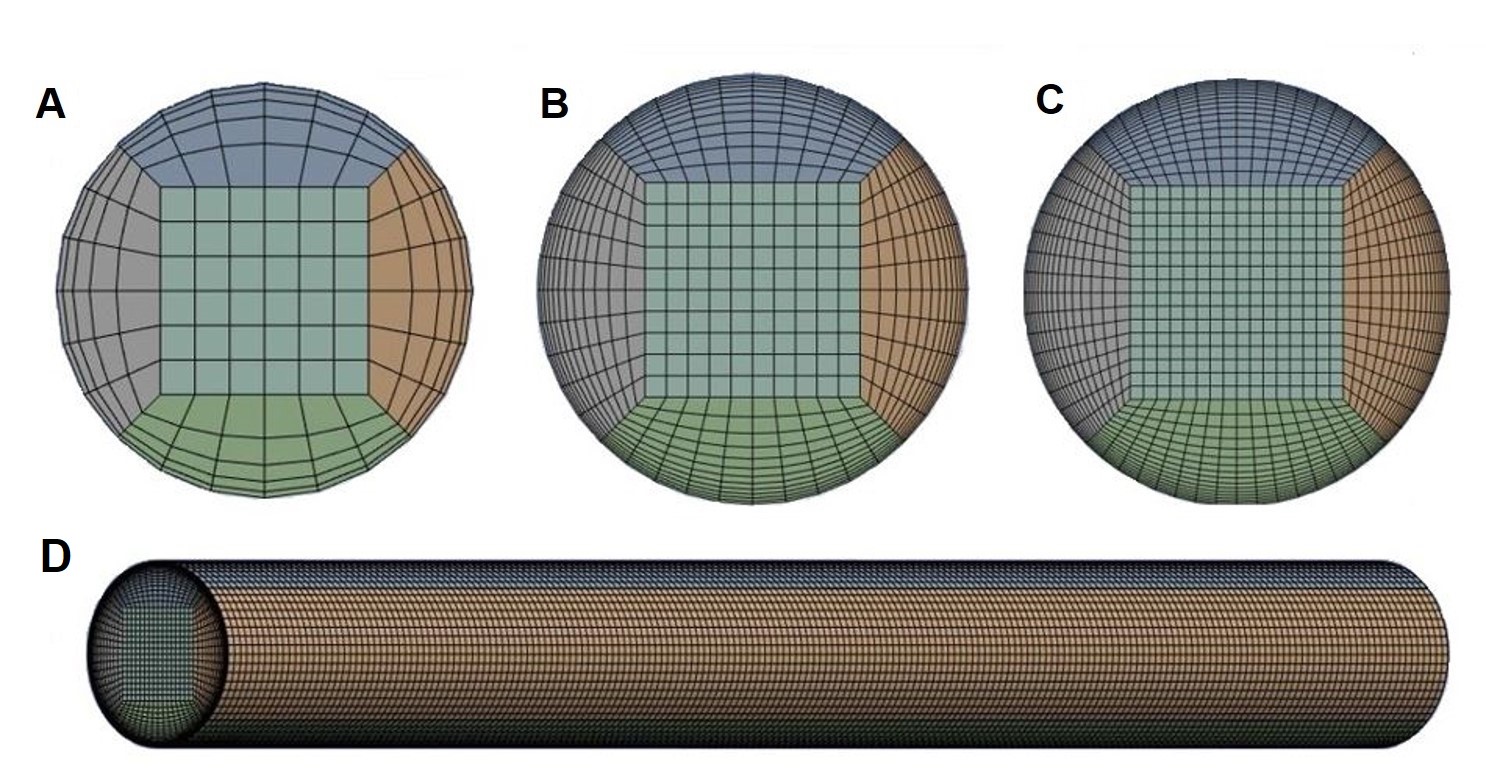} 
    \renewcommand{\captionfont}{\normalsize}
	\caption{ Computational model mesh structures (A) coarse mesh, (B) fine mesh, (C) extra fine mesh, and (D) 3D computational structure.}
	\label{f02}
\end{figure}

At first, the mesh density on flow characteristics is explored systematically. The study utilizes three different mesh configurations (Figure~\ref{f02}~A-C): coarse, fine, and extra fine mesh. The number of nodes for each mesh is as follows: 126,000 for the coarse mesh, 232,200 for the fine mesh, and 448,335 for the extra fine mesh. To ensure accurate computational modeling, the average mesh quality is assessed by examining the skewness, which ranges from 0.72 to 0.94 across all three meshes. The study also considers 40 boundary layers to accurately capture the near-wall effect. Figure~\ref{f02}D illustrates the 3D computational structured mesh employed in the study. Velocity profiles along the vertical reference line are analyzed at a specific location (Z = 2 $m$) for all cases.\\

\begin{figure}[!ht] 
	\centering
	\includegraphics[width=0.575\columnwidth]{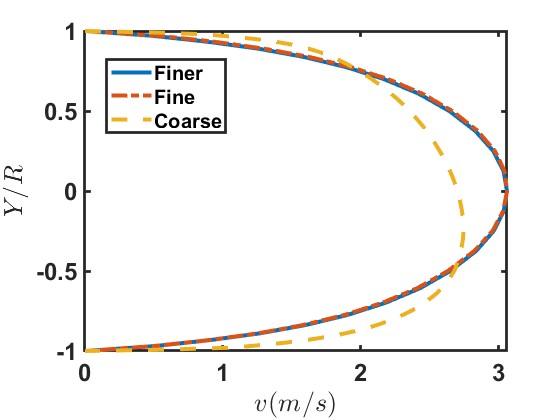} 
    \renewcommand{\captionfont}{\normalsize}
    	\caption{ Velocity profile for different mesh sizes for case 1. The corresponding operating conditions for case 1 are bitumen volume fraction of 0.0032, solid volume fraction of 0.1376, coal volume fraction of 0.0044, and 500$\mu$m bubble with the volume fraction of 0.0032  with a constant velocity of 2 $m/s$ at Z = 1.98 $m$.}
	\label{f03}
\end{figure}

The results presented in Figure~\ref{f03} demonstrate that the velocity profile is nearly identical between the fine and extra fine meshes. Therefore, it can be concluded that the fine grid, with its corresponding number of nodes, is sufficient for accurately capturing the flow physics. The maximum $y^{+}$ is 29.34. The cases used to investigate the bubble and coal influence on bitumen separation are given in Table~\ref{tab:Cases}.

\begin {table}[!ht]
\renewcommand\thetable{6}
\begin{center}
\caption { A list of cases and phases used in the CFD simulations.}
\label{tab:Cases}
\begin{tabular}{ l l l l l l }
\hline
case No. & bitumen vf &  solid vf &  coal vf &   bubble vf &  bubble size ($\mu$m) \\
\hline
case 1 & 0.0032 & 0.1376 & 0.0044 & 0.0032 & 500 \\
case 2 & 0.0032 & 0.1376 & 0.0044 & 0 &  \\
case 3 & 0.0032 & 0.1376 & 0 & 0.0032 & 500 \\
case 4 & 0.0032 & 0.1376 & 0 & 0 &  \\
case 5 & 0.0032 & 0.1376 & 0.0044 & 0.0064 & 500 \\
case 6 & 0.0032 & 0.1376 & 0.0044 & 0.0032 & 250 \\
case 7 & 0.0032 & 0.1376 & 0.0044 & 0.0032 & 100 \\

\hline

\end{tabular}
\end{center}

\end{table}
\subsection{Model Validation}


 To demonstrate the efficacy of E-E CFD model, the model's predictions are examined with the literature data for a highly concentrated  slurry transport in a horizontal pipeline~\cite{kaushal2005effect}. The inner diameter (D) of pipeline is 54.9 $mm$  and the length L= 3.3 m , and the slurry velocity $V$=~2 $m/s$ are considered for the model validation similar to the experimental conditions of ~\citet{kaushal2005effect}. As shown in Figure S2 at identical geometrical and operating conditions, the CFD model results are found to be excellent in agreement with experiments of concentration 20\% of \citet{kaushal2005effect}. Moreover, this validation establishes that the developed model is capable of forecasting the velocity and concentration distribution under experimental conditions. As a result, a well-validated model is accurate enough to predict flow behaviors within the range of operation conditions. In order to study bitumen distribution in slurry transport, the CFD model predictions are further extended to the conditions of in-house experiments. CFD simulations are primarily needed to understand how bitumen recovery is enhanced when coal is added, which will be discussed in the following section. 

\section{RESULTS AND DISCUSSION}

\subsection{Bitumen Removal with Separate Addition of Coal and Microbubbles }

\begin{figure}[!ht] 
	\centering
	\includegraphics[width=0.8\columnwidth]{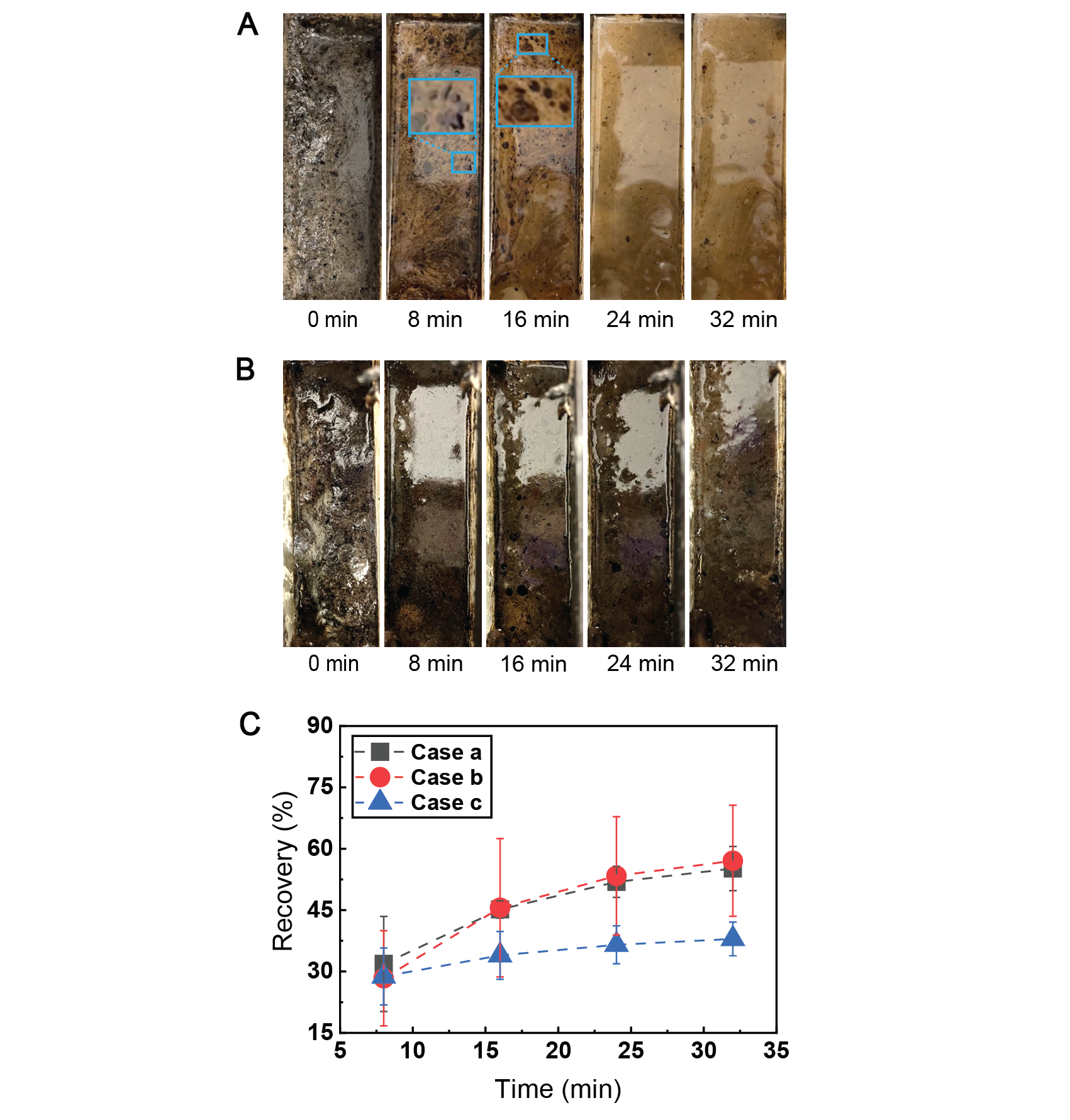} 
    \renewcommand{\captionfont}{\normalsize}
	\caption{ Froth observation in the trough at various intervals: (A) case a and (B) case c, and (C) bitumen recovery as a function of time under conditions of case a,b and c.}
	\label{f04}
\end{figure}

Figure \ref{f04} displays the appearance of recovered froth every 8 minutes in the recovering process. In absence of coal particles and $H_{2}O_{2}$ in the slurry, the amount of recovered froth decreased visually from 0 to 32 $min$ (Figure \ref{f04}A). At 0 $min$, the mixture of bitumen droplets and solids were clearly observed. The entrained gas immediately floated to the trough as bubbles that carry fine solids and bitumen after inserting into the loop.  Comparatively, there are many bitumen droplets but few solids at the surface at 8 and 16 $min$. Afterwards, dark bitumen droplets are infrequently spotted at 24 and 32 $min$, indicating that limited bitumen separation can be achieved spontaneously without the injection of microbubbles. Figure \ref{f04}B shows the observation of recovered froth with added coal particles of 40 $g$ at 5 intervals. At 0 min, the recovered froth with adding coal particles shows a darker appearance. From 8 min to 32 min, there is little difference in the froth appearance but the froth is darker in comparison to Figure \ref{f04}A. It might be due to the presence of floated coal particles in the trough.

The efficiency of bitumen recovery is shown in Figure \ref{f04}C as a function of time with various additives. An approximate 55 \% of bitumen residues were recovered at 32 $min$ with a slurry flow velocity of 2 $m/s$ and temperature of 42 \textdegree C. Interestingly, the recovery efficiency with the addition of microbubbles by $H_2O_2$ decomposition shares the similar increasing trend and a subtle increase in the cumulative recovery efficiency is observed at 32 $min$ in comparison to the recovery without additives. Even though microbubbles by $H_{2}O_{2}$ decomposition have shown an excellent performance in recovering bitumen \cite{zhou2023enhanced}, the volume of injected $H_{2}O_{2}$ and duration play a crucial role in microbubble generation. The low volume of $H_{2}O_{2}$ may not supply sufficient microbubbles required for interacting with the bitumen droplets. Simultaneously, the microbubble generation rate is too slow with the low concentration at a duration of 12 minutes.

Notably, there is little increment in the bitumen recovery from 8 to 32 $min$. The addition of coal particles achieved the bitumen recovery efficiency of 37.95 \%, indicating that coal particles hinders the spontaneous flotation of bitumen droplets. It affirms that the darker appearance of froth in Figure \ref{f04}B results from floated coal particles rather than bitumen droplets.
The hydrophobic surface of coal particles facilitates the hydrophobic interaction with bitumen droplets in the slurry. As reported in the study, bitumen suspensions can be removed by rigorous agitation in the presence of coal particles which can significantly enhance the bitumen-coal coagulation. However, the deposition of denser coal-bitumen aggregates hinders the flotation of bitumen droplets \cite{zhou2013role}.

\begin{figure}[!ht] 
	\centering
	\includegraphics[width=1\columnwidth]{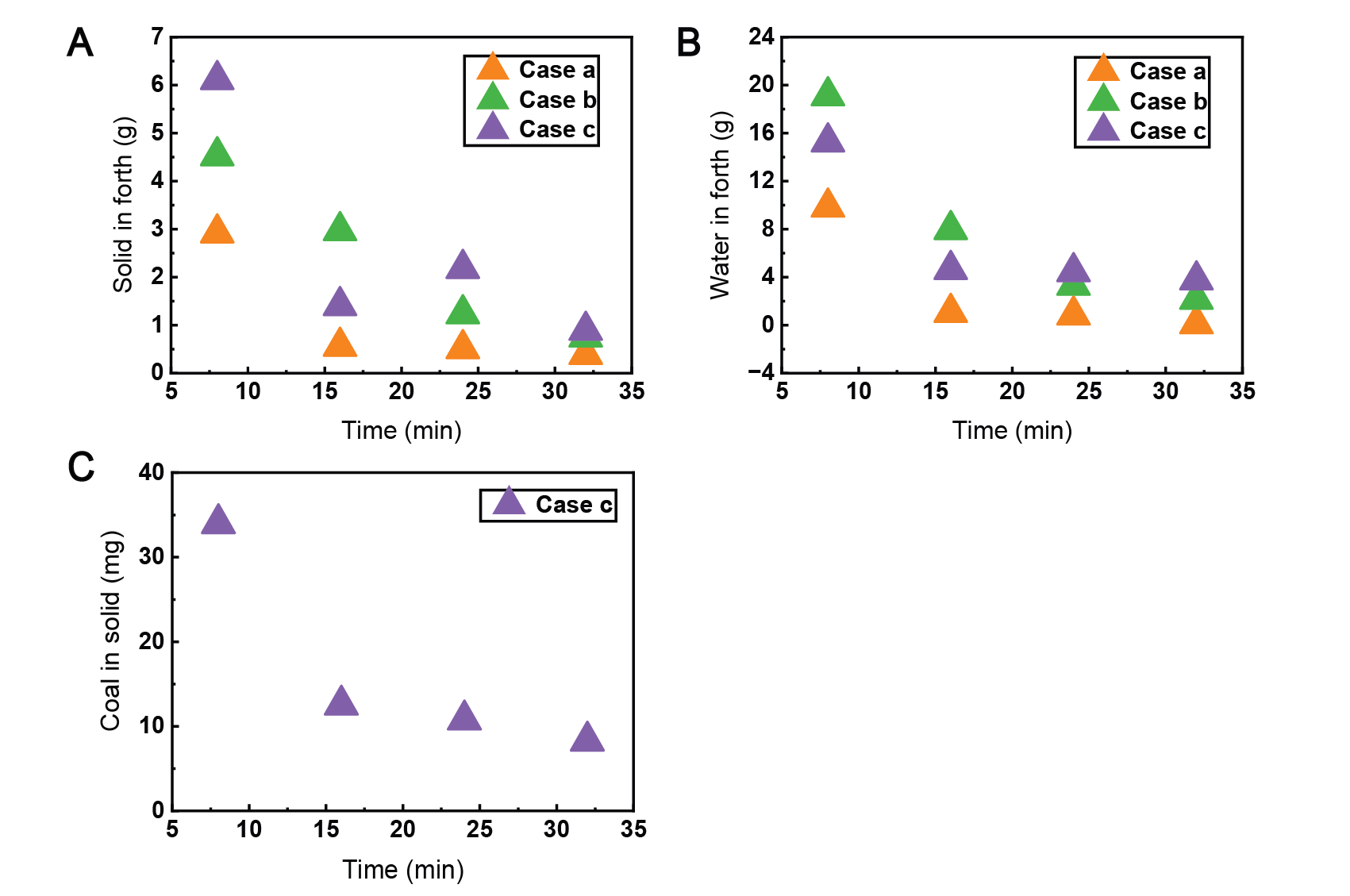} 
	\caption{Compositions of the froth in the blank case and cases where only coal or bubbles were added: (A) solid, (B) water, and (C) coal. }
	\label{f05}
\end{figure}

The water, solid, and coal contents in the froth are displayed in Figure \ref{f05}A-C. All the recovered contents at each interval decreased with time in general. The treatment without additives recovered the fewest amount of solids and water as seen in Figure \ref{f05}A and B. Typically, the highest amount of solids and water are recovered with coal particles and $H_{2}O_{2}$ separately. As a result, the coal particles only make up a small part of the recovered solids (Figure \ref{f05}C). It can be inferred that the floated coal particles did not contribute to the higher solid contents. Those tiny coal particles with rough surface might entrain more sands for flotation.

\subsection{Synergistic Effect of Microbubbles and Coal Particles on Bitumen Recovery}
\begin{figure}[!ht] 
	\centering
	\includegraphics[width=0.8\columnwidth]{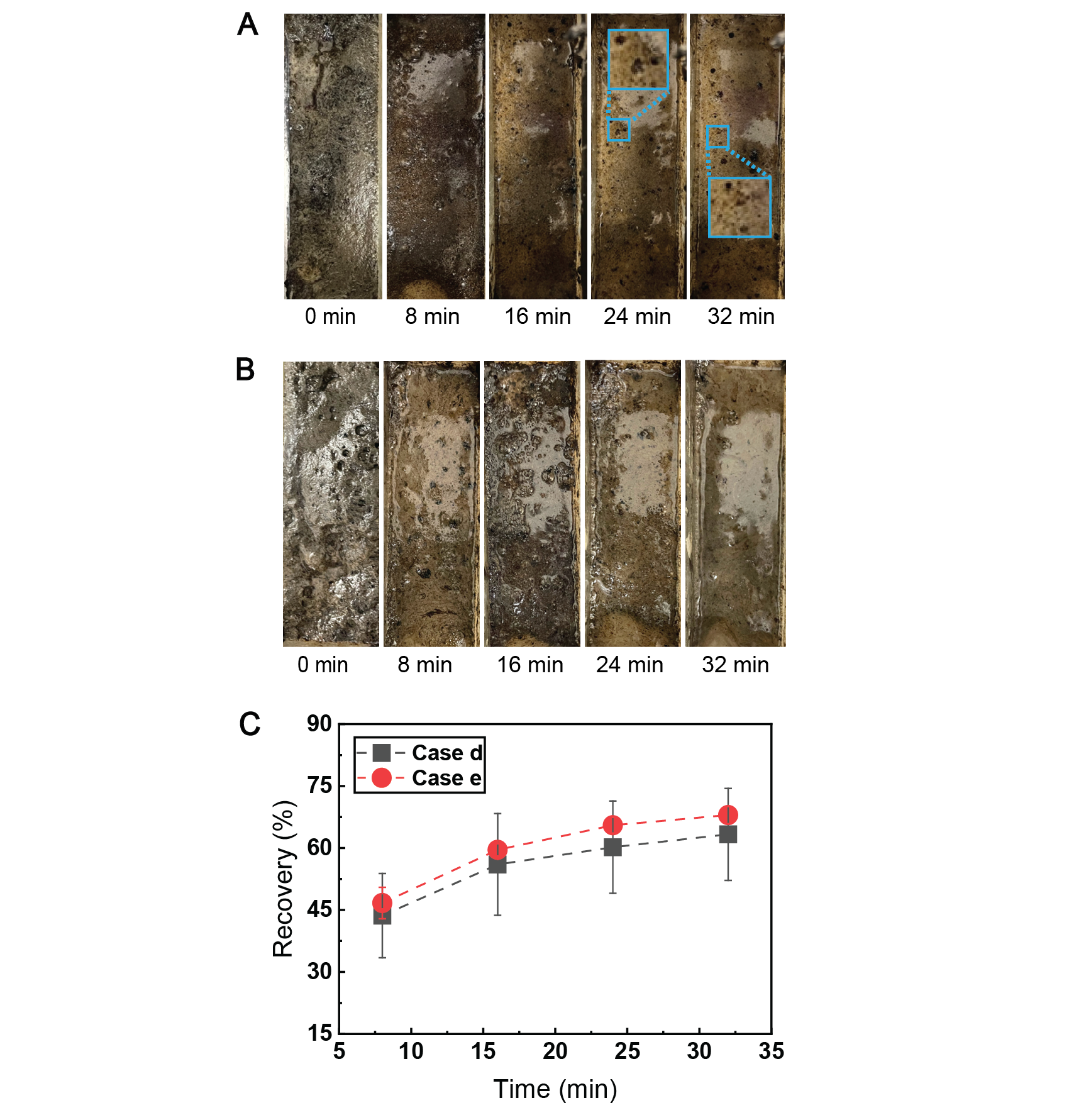} 
    \renewcommand{\captionfont}{\normalsize}
	\caption{Observation of froth in the trough at intervals: (A) case d and (B) case e, and (C) bitumen recovery as a function of time under conditions of case d and e.}
	\label{f06}
\end{figure}

We further investigate the simultaneous addition of coal particles and microbubbles into the oil sands tailings. In addition, the effect of injected volume of microbubbles is studied. Figure \ref{f06} depicts the observations of froth recovered with coal particles and $H_2O_2$ injection, respectively 6 and 12 $mL$. From Figure \ref{f06}A, it is evident that the froth composed of bubbles, bitumen, and solids are continuously separated from the tailings, but the froth color fades with time. It indicates that the contents of coal particles and bitumen in the froth might decrease with time. At 24 and 32 min, a few floating bitumen droplets without coverage of thick froth are visible at the froth surface. $H_2O_2$ injection accelerates the accumulation of froth containing dark bitumen droplets and coal particles at the trough. \\

As the injected volume of $H_2O_2$ was increased to 12 $mL$, the accumulated froth in the trough is thicker from 16 to 32 min (Figure \ref{f06}B) but fewer bitumen droplets are visible compared to that in Figure \ref{f06}A. It is certain that the higher volume of $H_2O_2$ produces a higher number of microbubbles that form a thicker froth layer composed of more solids. Those invisible bitumen droplets might be wrapped in the froth layer rather than float at the surface. As depicted in Figure \ref{f06}C, the cumulative recovery efficiency at 32 $min$ was enhanced slightly by 2 \% with the addition of 40 $g$ coal particles and injection of 6 $mL$ $H_2O_2$, while a significant increment of 12.8 \% was achieved in recovery efficiency when the amount of $H_2O_2$ was increased to 12 $mL$. For both injected volume of $H_2O_2$, a rapid increase of recovery efficiency can be seen in the first 16 minutes and the growth trend of recovery efficiency flattens since 16 $min$. It is interesting to find that the trend of bitumen recovery with time was greatly associated with the $H_2O_2$ injection. The microbubble generation rate depends on the decomposition rate and concentration.~\citet{hiroki2005decomposition} found that the decomposition rate decreases with the increase of $H_2O_2$ concentration due to the higher activation energy. In our previous study, the visualization of formed bubbles in the pipeline showed that the number of formed bubbles increased as the injection of $H_2O_2$ but rapidly decreased at 32 min after the injection\cite{zhou2023enhanced}. This phenomenon demonstrates that the $H_2O_2$ concentration plays a dominant role in generating bubbles although the decomposition rate might be improved at a lower concentration. The higher $H_2O_2$ concentration produces a higher number of microbubbles. The microbubble generation might decay shortly after the injection, resulting in decrease in the receovry rate for a shorter injection time. It agrees well with the observation in this study that the froth content decreased apparently after 16 min with the injection time of 12 minutes. \\

In bitumen flotation, the high density of coal particles is an adverse factor. Hydrophobic interaction between bitumen droplets and coal particles increases the size of bitumen droplets that enhances the attachment to bubble surface. On the other hand, the attachment probability can be greatly improved by the larger number of microbubbles at a higher $H_{2}O_{2}$ concentration.\\

\begin{figure}[!ht] 
	\centering
	\includegraphics[width=1\columnwidth]{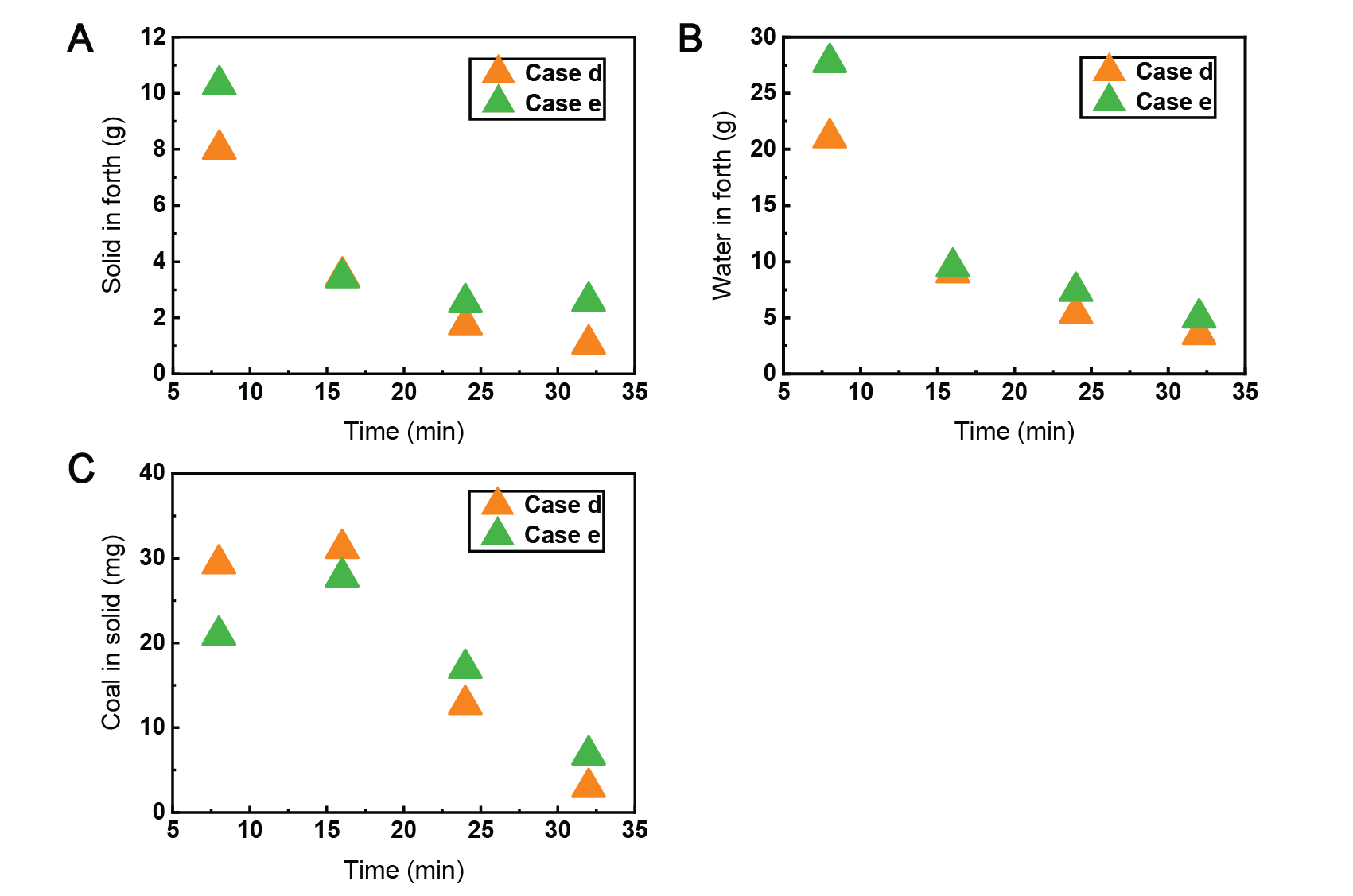} 
    \renewcommand{\captionfont}{\normalsize}
	\caption{Compositions of the froth in the cases with the same amount of coal added but different amounts of bubbles: (A) solid, (B) water, and (C) coal. }
	\label{f07}
\end{figure}

Figure~\ref{f07}A-C show the water, solid and coal contents in the froth respectively. It should be noted that both water and solid contents decrease with time, and a higher amount of water and solids are recovered with 12 $mL$ $H_2O_2$ injection. The higher number of microbubbles can carry more solids and water in the froth. The collected coal contents experienced a increase at the first two collections and then a sharp decrease afterwards. Additionally, the higher  volume of injected $H_2O_2$ did not lead to more recovery of coal particles. It is interesting that the recovering trend of coal particles and bitumen are similar with the increase in injection of $H_2O_2$. The pronounced bubble generation during injection might enhance the coal recovery at 8 and 16 min. Those large particles first to be collected by the bubbles to the froth. Moreover, there were fewer small coal particles and bitumen droplets remained in the tailings after the tremendous recovery. It also contributed to the decreased interaction between bubbles and coal particles. 

\subsection{Simulation Results}
\subsubsection{The Addition of Coal}

\begin{figure}[!ht] 
	\centering
	\includegraphics[width=1\columnwidth]{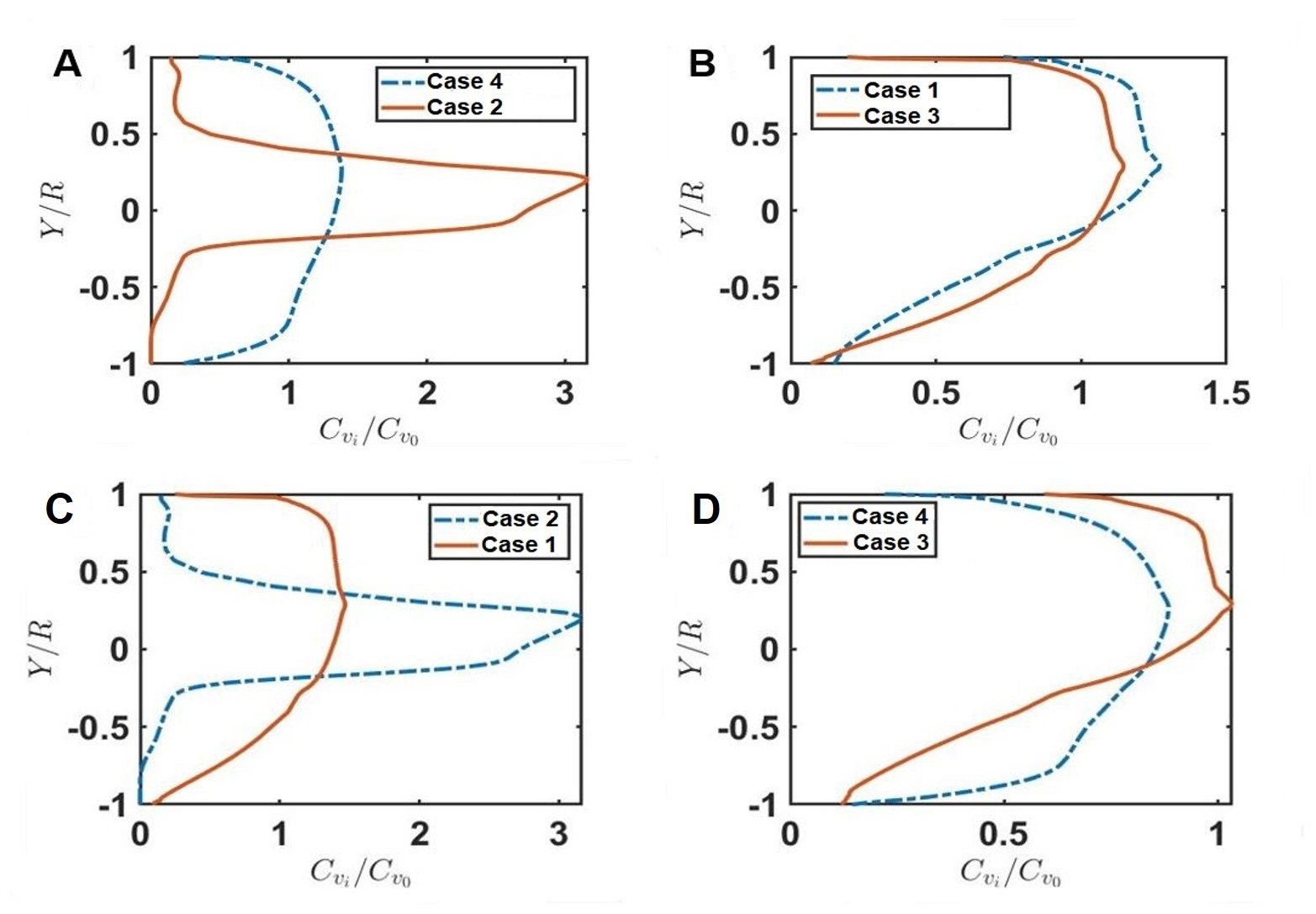} 
    \renewcommand{\captionfont}{\normalsize}
    \caption{ Bubbles influence on bitumen distribution: (A) Coal impact without bubbles - bitumen accumulates mid-section with coal (case 2), and floats at the top without coal (case 4). (B) Coal effect with bubbles - bubble-bitumen interaction shifts mid-section concentration to the pipe's crown (case 1), and bubble bitumen interaction without coal causes less accumulation at pipe crown (case 3). (C) Bubble impact with coal - bubbles alter bitumen movement, accumulating at the top (case 1), coal impact with out bubble leads to bitumen accumulation in mid-section (case 2). (D) Bubble effect without coal- bubbles cause bitumen accumulation at the top section (case 3), and the bitumen accumulation without bubble is less (case 4)}
	\label{f10}
\end{figure}

This section illustrates the impact of coal particles on bitumen recovery. The coal particles are hydrophobic and possess a lower density compared to the solid particles, although they are still more dense than the carrier fluid. This property causes the coal particles to occupy the mid-section of the pipe, affecting the bitumen distribution along the pipeline to be enchained in the mid section of the pipeline when there is no bubble (Figure \ref{f10}A). In the absence of bubbles in the system, this mid-section acts as a blockade, preventing bitumen droplets from accumulating at the top section of the pipeline. Simultaneously, the zone causes the bitumen to be trapped along the small solid particles, forming chains that do not surround the solid bed. Consequently, when only coal is present in the system, the bitumen recovery rate decreases on the top section of the pipeline.\\

However, when bubbles are introduced into the system, the interaction between droplets and bubbles significantly affects the bitumen trapped in the coal particle zone. This interaction liberates the bitumen droplets, allowing them to accumulate at the upper portion of the tube due to the enchained bitumen at the midsection the bubbles will have more interactions with bitumen and therefore the bitumen accumulation at the pipeline's crown will be greater than the case when the bubble interact with bitumen through a slurry without coals (Figure \ref{f10}B). As some bitumen droplets are removed from this section due to the droplet/bubble interaction, the settled solids in this zone also entrap some bitumen droplets within the solid bed at the pipeline's invert section. To elucidate the influence of coal particles in the system, the study examines the behavior of coal particles in the presence and absence of these particles in the slurry flow. The results are analyzed after 28 seconds and at a length of 1.98 meters, representing a stable state of the velocity profile and a fully developed flow condition.

\subsection{The Addition of Bubbles}

Bubbles play a critical role in oil-sand systems to facilitate the separation of bitumen from solid content in the slurry. The interaction between bitumen droplets and bubbles results in the attachment of bitumen droplets to the bubbles. Due to the lower density possessed by both the bubble and bitumen compared to the carrier fluid. The bitumen accumulates at the top section of the pipeline, effectively separating it from the solid content. The presence of tiny bubbles ensures a larger surface area, increasing the likelihood of bubbles attaching to bitumen droplets. To comprehend the influence of bubbles on bitumen recovery in the slurry flow, the volume fraction of bubbles is considered equal to the volume fraction of bitumen. The results are analyzed after 28 seconds and at a length of 1.98 meters, representing a stable state of the velocity profile and a fully developed flow condition. The bubble effect is illustrated in Figure \ref{f10}C, in which bitumen enchained in the center will be released and carried to the pipe's top section by bubbles. Finally, the bubbles effect in a slurry without coal has been demonstrated (Figure \ref{f10}D), which would cause bitumen accumulation on the top section of the pipeline.   \\

\subsection{Effect of Bubble Size}

\begin{figure}[!ht] 
	\centering
	\includegraphics[width=1\columnwidth]{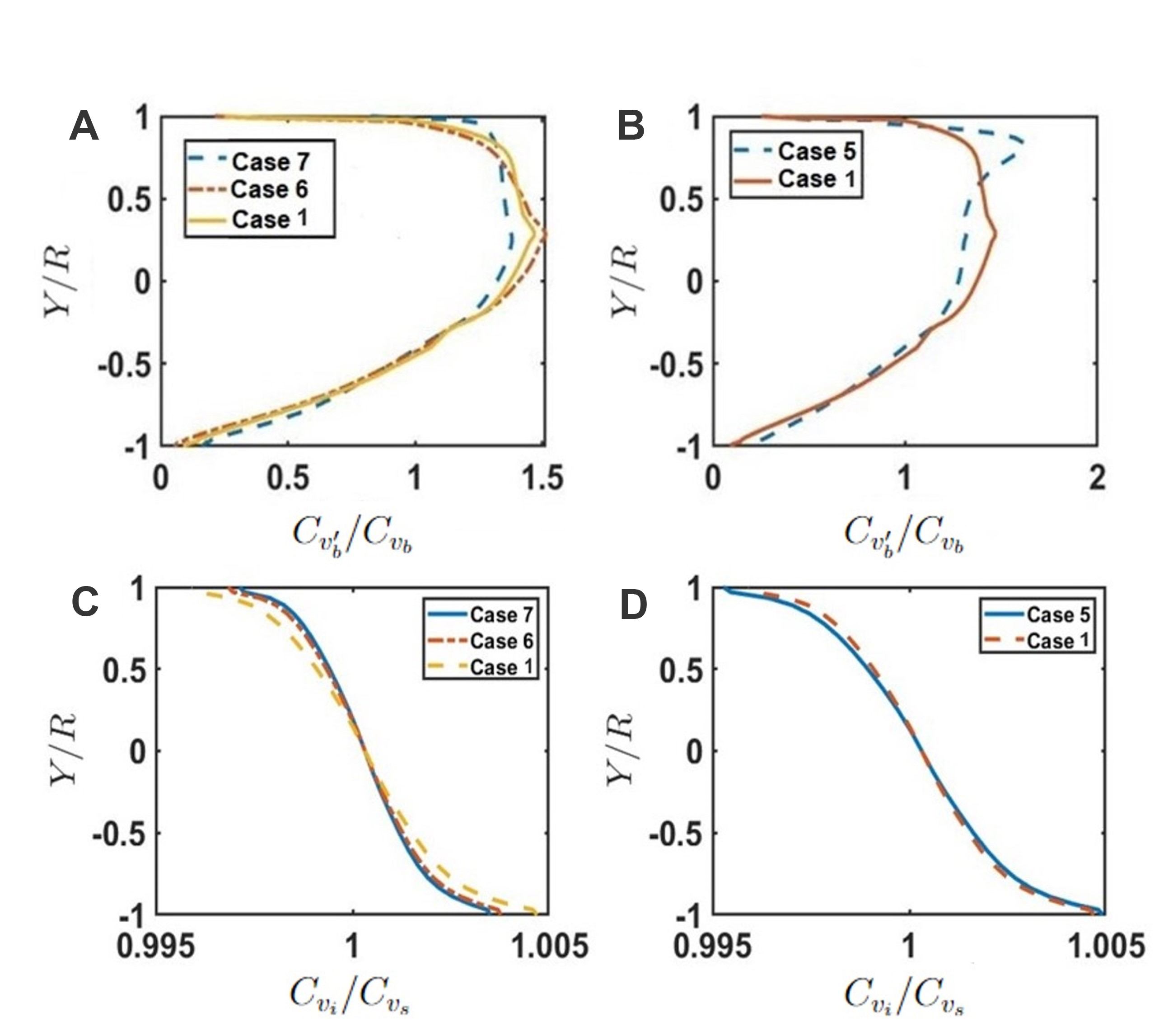} 
    \renewcommand{\captionfont}{\normalsize}
	\caption{Bubble size and volume fraction impact: (A) bitumen distribution influenced by bubble size, (B) bitumen distribution influenced by bubble volume fraction, (C) solid distribution influenced by bubble size, and (D) solid distribution influenced by bubble volume fraction  (VF: Volume Fraction)}
	\label{f11}
\end{figure}

Bitumen separation relies heavily on bubble size, since an increase in bubble size will result in an increase in bubble volume, thus increasing bubble bounce. The greater the bouncy force would indicate a greater driving force for the movement of the bubbles to the pipeline crown. Moreover the larger bubbles would have a greater momentum, due to this phenomenon the bubble-bitumen interaction would have a greater impact though larger portions of bitumen would be separated and accumulated at the top section of the pipeline. As illustrated in Figure~\ref{f11}A, various bubble sizes were investigated in case 1, case 6 and case 7 while maintaining a constant volume fraction. Larger bubbles possess a greater volume and, consequently, a higher buoyant force. This buoyant force leads to increased momentum, causing the bubbles to accumulate more rapidly at the top section of the pipeline. However, as discussed in the previous section, the volume fraction is a constant parameter when investigating the bubble size influence, larger bubbles result in fewer bubbles and a reduced interacting surface area compared to smaller bubbles leading to fewer bitumen droplet/bubble interactions. 

The considered bubble sizes for case 7, case 6 and case 1 are 100, 250, and 500 $\mu$m, respectively. An optimal condition exists where bubbles are large enough to facilitate faster accumulation at the top while remaining small enough to efficiently interact with bitumen droplets. However, while bubbles accumulate along the top section of the pipeline, the bitumen profile reveals a gradual increase in accumulation along the pipeline length. The interaction of bubbles with bitumen droplets results in the detachment of bitumen from solid and coal particles. Additionally, the occupation of bubbles along the bitumen at the pipe's crown causes solids and coals to be unoccupied from the pipeline crown and the solid content would slightly decrease form the pipeline crown and slightly increase at pipeline invert, In case 1 the larger bubbles resulted in a larger accumulation of bitumen at top of the pipeline while the solid profile also shows that more solid had been unoccupied from the pipeline crown and moved to the invert compared to case 6 corresponding to bubble size of 250~$\mu$m as well as case 7 with bubble size of 100~$\mu$m, the same outcome can be seen comparing case 6 and case 7, where the larger bubbles of case 6 had the same outcome compare to case 7 (Figure \ref{f11}C). The results were analyzed after 28 $s$ and at a length of 1.98 $m$ when the velocity profile attains a stable state and reaches a condition of fully developed flow profile. 


\subsection{Effect of Bubble Volume Fraction} 

\begin{figure}[!ht] 
	\centering
	\includegraphics[width=0.95\columnwidth]{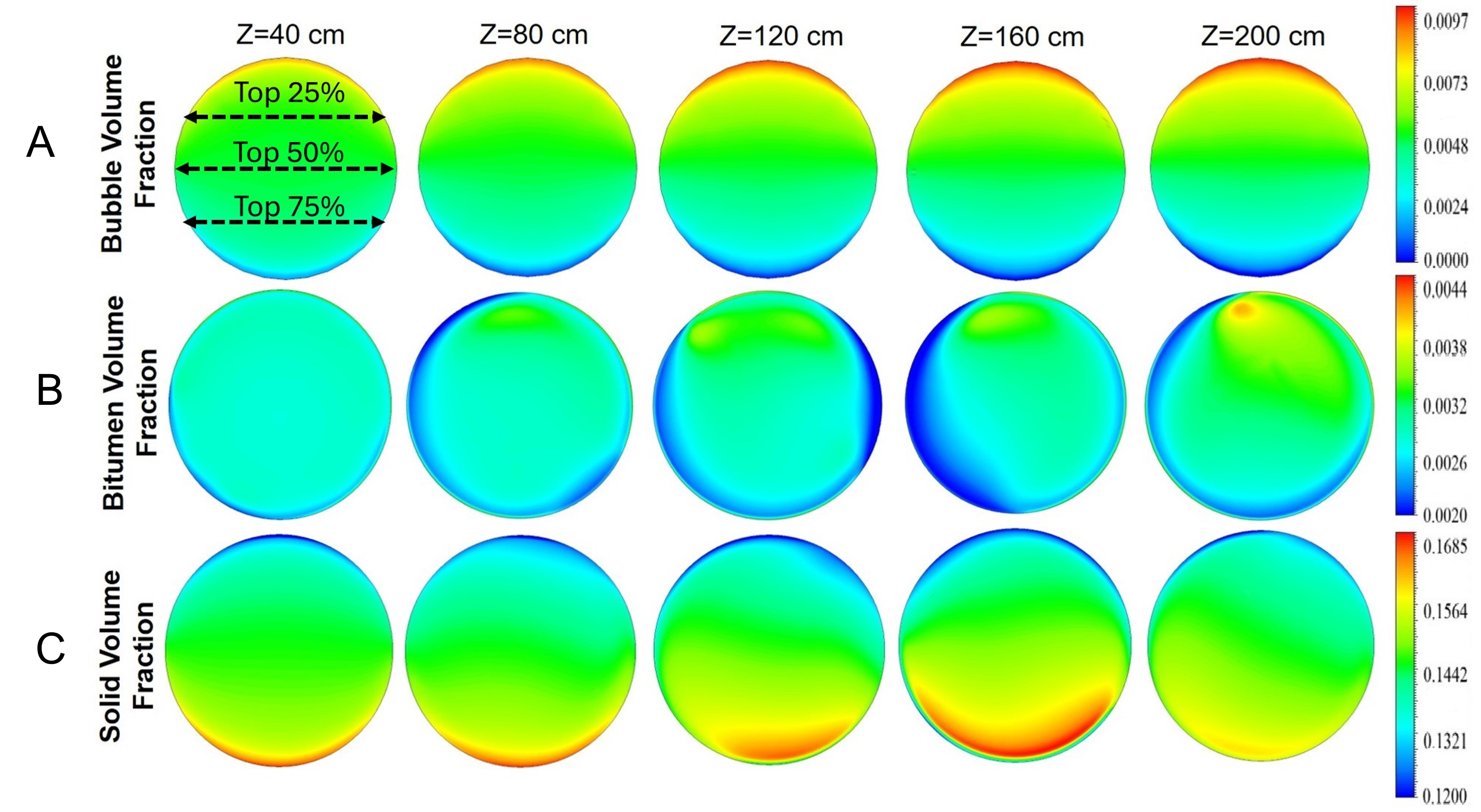} 
    \renewcommand{\captionfont}{\normalsize}
	\caption{The slurry content distribution over the pipe's cross section at different lengths from inlet for Case 1 : (A) Bubble distribution contour has been demonstrated indicating the bubble accumulation on pipe's crown, (B) bitumen distribution contour shows that the bitumen volume fraction will increase on top section of the pipe, (C) solid distribution profile reveals the solid settling at pipe's invert.}
	\label{f12}
\end{figure}

    To examine the influence of bubble volume fraction, two cases have been considered, case 1 and case 5, which respectively have the bubble volume fraction of 0.0032 and 0.0064, while in both cases the bubble size is 500 $\mu$m. The increase of bubble volume fraction in case 5 compared to case 1 leads to greater number of bubbles. The interaction between the bubbles and bitumen droplets would increase which would lead to enhanced separation of bitumen from solid content and an increased accumulation of bitumen at the pipeline's crown as depicted in Figure \ref{f11}B. The normalized bitumen concentration profiles were analyzed with the dimensionless quantity $C_{v_b^{\prime}} / C_{v_b}$, where $C_{v_b^{\prime}}$ and $C_{v_b}$ represent the local concentration of bitumen and the bulk concentration of bitumen as shown in Figure \ref{f11}C. As the pipeline crown is occupied by greater number of bubbles and bitumen droplets more solid particles would be unoccupied from the pipeline crown and their concentration would in crease in the pipeline invert. Though due to small volume fraction of bitumen and bubble compared to the solid volume fraction, the solid profile would change slightly (Figure \ref{f11}D). The normalized concentration profiles compared for difference  cases with dimensionless quantity $C_{v_i} / C_{v_s}$, where the  $C_{v_i}$ and $C_{v_s}$ are the local concentration of solids on a specific $i^{th}$ location and bulk concentration of solids. Two cases were studied, with bubble volume fractions equal to the bitumen volume fraction and twice the bitumen volume fraction. The results were examined after 28 $s$ and at a length of 1.98 $m$, corresponding to stable and fully developed flow profile conditions of the velocity profile. Table~\ref{tab:bitumen-bubble} presents the bitumen recovery summary, taking into account variables such as bubble size, bubble fraction, and the presence or absence of coal and bubbles. Considering the diameter of the pipeline from the invert to the crown of the pipeline, the surface area corresponding to the top quarter of the pipeline vertical diameter is considered the top 25$\%$, from middle to top of the pipeline is considered top 50$\%$, and the area corresponding area from the top 75$\%$ of pipeline are considered for the analysis as shown in Figure \ref{f12}A. The bitumen accumulation for different cases are demonstrated in Table \ref{tab:bitumen-bubble}. \\


\renewcommand{\arraystretch}{1}
\renewcommand\thetable{7}
\begin{centering}
\begin{table}[!ht]
  \caption{Bitumen recovery and bubble characteristics.}
\label{tab:bitumen-bubble} 

\begin{tabular}{ l l l l l l l }

\hline

bitumen recovery &  &  & & top (25$\%$) & top (50$\%$) & top (75$\%$)\\
\hline
& \centering bubble Size ($\mu$m) & \centering bubble vf & \centering coal vf  & & &  \\
\hline
case 1 & \centering 500 & \centering 0.0032 & 0.0044 & 30.1 $\%$ & 62.4 $\%$ & 89.2 $\%$ \\
case 2 & \centering --- & \centering --- & 0.0044 & 11.4  $\%$ & 67.9 $\%$ & 95.1 $\%$ \\
case 3 & \centering 500 & \centering 0.0032 & --- & 28.1 $\%$ & 60.6 $\%$ & 87.6 $\%$ \\
case 4 & \centering --- & \centering --- & --- & 24.1 $\%$ & 54.0 $\%$ & 80.8 $\%$ \\
    case 5 & \centering 500 & \centering 0.0064 & 0.0044 & 33.8 $\%$ & 64.0 $\%$ & 91.1 $\%$ \\
case 6 & \centering 250 & \centering 0.0032 & 0.0044 & 29.5 $\%$ & 61.6 $\%$ & 88.6 $\%$ \\
case 7 & \centering 100 & \centering 0.0032 & 0.0044 & 29.7 $\%$ & 60.7 $\%$ & 87.0 $\%$ \\

\end{tabular}

\end{table}
\end{centering}

The coal particles in a mixture when the bubble does not exist cause the bitumen to be enchained and trapped mostly at the mixture without accumulating at the top or having a homogeneous distribution, as the bubbles are introduced to the system the coal-bitumen interaction is replaced by the bubble-bitumen interaction and while the bubbles liberate the trapped bitumen they will accumulate on top. Moreover, the bitumen droplets would be dragged to the top section of the pipeline due to bubble-bitumen interaction. As a result, the bitumen volume fraction on the top would be greater, also if the coals and bubbles are presented in the mixture (i.e., cases 1, 3, 5, 6, and 7) compared to cases in which there are only coal (i.e., case 2) the accumulated bitumen is found to be maximum in the top 75\% of the pipe cross-section (Table \ref{tab:bitumen-bubble}). The bitumen distribution profile indicates that most of the bitumen droplets are at the top surface of the pipeline (Z = 160 cm) as shown in figure \ref{f12} A. As a result minimum bitumen droplets are found near the wall. Moreover, the increase of bubble size can increase the buoyancy force and a greater force to move the bubble to the pipe's crown therefore a greater accumulation which leads to more accumulation of bitumen, the number of bubbles would be increased by an increase in volume fraction while the bubble size is constant, the more number of bubbles would lead to more interactions between the bubbles and bitumen and therefore an increased accumulation of bitumen occurs at the upper intersection of the pipe. Eventually, solid particles would settle at the bottom of the pipeline as shown in Figure \ref{f12}C.  \\

The volume fraction contours depicted in Figure \ref{f12} illustrate the accumulation of bubbles (Figure \ref{f12}A), bitumen (Figure \ref{f12}B), and solid particles (Figure \ref{f12}C) across the cross-section of the pipeline at various lengths (i.e., Z = 40 cm, 80 cm, 120 cm, 160 cm, and 200 cm) from the inlet to the outlet for case 1. Bubbles accumulate along the upper part of the pipeline as it progresses from the inlet to the outlet. Additionally, interactions between the bubbles lead to the accumulation of bitumen at the crown of the pipeline. Moreover, interactions between coal and bitumen cause the bitumen to concentrate in the central section of the pipeline. A notable shift in the distribution of bubbles occurs when a majority of them accumulate at the crown of the pipeline (Figure \ref{f12}A at 160 cm and Figure \ref{f12}A at 200 cm). As the pipeline's midpoint is reached, the interaction between coal and bitumen begins to predominate, while the interaction between bitumen and bubbles diminishes. Consequently, the accumulation of bitumen slows down, and the distribution of bitumen is primarily influenced by the interaction between coal and bitumen at the interface of the pipeline (Figure \ref{f12}B at 160 cm and Figure \ref{f12}B at 200 cm).

\newpage
\section{Conclusions}
In this study, we explored a novel method for recovering bitumen residues from highly concentrated oil sands tailings, utilizing microbubble injection enhanced by coal particles. Our comprehensive analysis, encompassing both experimental and simulation results, sheds light on the intricate dynamics of this process.
Our findings revealed an interesting phenomenon: the presence of coal particles appears to hinder the spontaneous recovery of bitumen. While simulation suggested enhanced flotation with microbubbles, the experimental results showed only marginal increases in bitumen recovery following microbubble injection. To delve deeper into the role of microbubbles, our CFD simulations highlighted two critical factors: increasing bubble size and increasing the volume of injected bubbles can significantly facilitate bitumen flotation.
However, the real breakthrough came when microbubbles and coal particles synergized. The combined effect led to a substantial enhancement in recovery efficiency. Furthermore, extending the duration of microbubble injection proved to be a valuable strategy, resulting in a notable increase in cumulative bitumen recovery over time.
In summary, our study provides essential insights into the role of fine hydrophobic particles in bitumen separation facilitated by bubble flotation. These findings can be extended to the separation of other hydrophobic minerals and organic matter, offering tremendous potential for environmental waste removal and boosting production in various engineering applications. The scalability, cost-effectiveness, and process optimization are challenges that still need to be addressed in future research.\\

To enhance its practical significance, the systamatic investigations on various operating conditions, such as different injection rates of coal particles and hydrogen peroxide, will maximize bitumen recovery while minimizing energy consumption and operational costs. In addition, the CFD simulations will enable us to comprehensively understand the complex dynamics of the system and guide the design of more efficient and scalable industrial processes.

\newpage
\section*{Data availability}

The data that support the findings of this study are available from the corresponding author upon reasonable request.

\begin{suppinfo}

A supporting document is available on the kinetic theory of granular flow model equations for solid and coal phases; and the E-E CFD model solver settings.
\end{suppinfo}
\section*{Credit author statement}
Y. Huo, M. Golchin, and K. Zhou contributed equally to this work. \textbf{Y. Huo:} Conducting experiment, Methodology, Formal Analysis, Writing – Original Draft, Writing – review \& editing. \textbf{K. Zhou:} Conducting experiment, Methodology, Formal Analysis, Writing – Original Draft, Writing – review \& editing. \textbf{M. Golchin:}  CFD simulations, Software, Methodology, Formal Analysis, Writing – Original Draft, Writing – review \& editing. \textbf{A. Abraham:} Conducting experiments. \textbf{S.G Sontti:} CFD simulations, Model validation, Methodology, software, visualizations, Writing – review \& editing. \textbf{X. Zhang:} Conceptualization, Project administration, Writing – review \& editing, Resources, Supervision.
\section*{Declaration of competing interest}
 The authors declare that they have no known competing financial interests or personal relationships that could have appeared to influence the work reported in this paper.


\begin{acknowledgement}

The authors are thankful to the support from Canada Research Chairs Program, Discovery Project, Alliance Grant from Natural Sciences and Engineering Research Council of Canada. We also thank the Digital Research Alliance of Canada (https://alliancecan.ca/en) for continued support through regular access of high-performing server HPC cedar and Graham Cluster. 
 This work was supported by Imperial Oil Limited and Alberta Innovates through the Institute for Oil Sands Innovation at the University of Alberta (IOSI). This work was also supported by the China Scholarship Council for Y. Huo (202206430010), and K. Zhou (202008180018).

\end{acknowledgement}


\section*{Nomenclature}

\begin{longtable}{l p{12cm}}
$D$ &  pipe diameter (L) \\
$R$ & pipe radius (L)\\
$d_p$ & particle diameter (L)\\
$C_{v'_{b}}$ & chord\textendash concentration of bitumen (--)\\
$C_{v}$ & chord\textendash averaged concentration (--)\\
$C_{v_{i}}$ & local concentration (--)\\
$C_{v_{s}}$ & bulk concentration of solids(--)\\
$g$ & gravitational acceleration (L T$^{-2}$)\\
$g_0$ & radial distribution function (--)\\
$p$ & locally\textendash averaged pressure (M L$^{-1}$ T$^{-2}$)\\
$t$ & time (T) \\
$v$ & velocity (L T$^{-1}$)\\
$V$ & velocity (L T$^{-1}$)\\
$F$ & force per unit volume (M L$^{-2}$ T$^{-2}$)\\
$F_{l,si}$ & interaction force between the liquid and $i^{th}$ solid  phase (M L$^{-2}$ T$^{-2}$) \\
$C_\mathrm{fr}$ & friction coefficient between solid phases (--)\\
$x$ & horizontal coordinate (L)\\
$y$ & vertical coordinate (L)\\
$z$ & axial coordinate (L)\\
$e$ & restitution coefficient (--)\\
$k$& turbulent kinetic energy (L$^{2}$ T$^{-2}$)) \\ 
$G_{k,m}$ & turbulence generation rate (M L$^{-1}$ T$^{-3}$)	\\ \\

	\textit{Greek symbol}\\ 
	
$\alpha$ & locally\textendash averaged volume fraction (--)\\
$\mu$ & dynamic viscosity (M L$^{-1}$ T$^{-1}$)\\
$\rho$ & density (M L$^{-3}$)\\
$\phi_{ls}$ & the energy exchange between the fluid and the solid phases (E) \\
$\gamma_{\Theta_{s}}$ & collisional dissipation of energy (E)\\
$\tau$ & shear stress (M L$^{-1}$ T$^{-2}$)\\
$\dot{\gamma}$ & shear strain rate (T$^{-1}$)\\
$\alpha_{s,\mathrm{max}}$ & maximum packing limit (--)\\
$\Theta$  & granular temperature (L$^{-2}$ T$^{-2}$)\\
$\varphi$ & angle of internal friction (--)\\
$\mu_{t,m}$ & turbulence viscosity (M L$^{-1}$ T$^{-1}$)\\ 
$\epsilon$ & dissipation rate  (L$^{2}$ T$^{-3}$)\\ 
$\lambda$ & bulk viscosity (M L$^{-1}$ T$^{-1}$) \\
$\rho_{m}$ & mixture density \\ \\

\textit{Subscripts}\\
	
	$l$ & liquid\\
	$s$ & solid\\
    $c$ & coal\\
    $b$ & bitumen\\
	$si$ & solid $i^{th}$ phase \\
	$i$ & $i^{th}$ solid phase\\
    $j$ & $j^{th}$ solid phase\\
	$q$  & $q^{th}$ solid phase\\
	col & collisional part of viscosity\\
	 kin & kinetic part of viscosity\\
	fr & frictional part of viscosity  \\ 
	td & turbulent dispersion \\
    drag &  drag \\
    0 & initial \\
\end{longtable}
\cleardoublepage
\bibliography{ref}

\end{document}